# Comment on "Gold Nanobipyramid-Enhanced Hydrogen Sensing with Plasmon Red Shifts Reaching ≈140 nm at 2 vol% Hydrogen Concentration"

Aradhna Jeevan[1,2,*] and Yexing Ma[2,**]

[1]Centre for Chemical Sciences, School of Basic and Applied Sciences Central University of Punjab, Bathinda, India

[2]School of Materials Science and Engineering, University of New South Wales, NSW, Australia

[*] Corresponding Author: jeevanka@yahoo.com

[**] Corresponding Author: yexingma@hotmail.com

**Summary:** In the recently-published paper "Gold Nanobipyramid-Enhanced Hydrogen Sensing with Plasmon Red Shifts Reaching ≈140 nm at 2 vol% Hydrogen Concentration" by Yip et al.,[1] the authors claimed that the catalytic activities of the synthesized Au nanobipyramids(NBP)@Pd nanostructures could be mainly due to the electronic effect. Here, we show that this claim is unsupported and an important reference was mis-cited in this commented publication.

To understand the alloying effects for adsorption, there are at least three types of effects that should be clarified: ensemble, strain and electronic effects. The ensemble effects are the various arrangements of atoms on the adsorption sites, which are recently considered as the most important effects for catalysis on randomly-mixed alloy structures. The strain effects originate from the mismatch of the lattice constants of different types of alloyed elements, leading to significant differences on the surface bond length. However, compared to the ensemble and strain effects, the electronic effects influence more slightly to the adsorption for an alloyed nanostructure. There are recently plenty of studies that have evaluated these effects from both experiments and theories.[2–9]

However, in the recently-published paper "Gold Nanobipyramid-Enhanced Hydrogen Sensing with Plasmon Red Shifts Reaching ≈140 nm at 2 vol% Hydrogen Concentration" by Yip et al.,[1] the authors claimed that the hydrogen sensing of the synthesized Au nanobipyramids(NBP)@Pd nanostructures was mainly due to the electronic effect. Two references were cited in order to support such an explanation. However, in one of the references cited by the paper (Ref.[4] by Luo et al.), it actually proves the opposite conclusion: ensemble effects should be the most dominate effects for an alloy system in the cases of PdAu and PtAu nanoparticles (NPs). Both experiments and theories have proven such a phenomenon, as shown in Ref. [4]. Using both electrochemical experiments and density functional theory (DFT) calculation techniques, Luo et al. also have shown that H adsorption is extremely sensitive to the adsorption environment on alloyed catalytic surfaces. Compared to ensemble effects, electronic effects on alloyed systems are much less significant, as proven by many studies from both experimental and theoretical techniques.[3–6,8,9] Also, Pd and Au have good miscibility.[10] This leads to the high possibility that there occurs surface rearrangement of Au on PdAu core@shell structures, forming alloyed surface with various Pd-Au ensembles.[11] In the authors' characterizations, there is no evidence that could exclude this possibility. Therefore, the commented paper published by Yip et al. mistakenly cited the paper that concluded for ensemble effects and claimed that electronic effects are the most dominant in their catalytic systems, which is misleading to readers from catalysis background.

As for the strain effect, due to the significant difference of lattice constant between Au and Pd, it is expected that the Pd-Pd bond lengths would be expanded with the alloying of Au. Such a change in bond length, would also dramatically tune the adsorption energy during catalysis. Similar studies done by Zhang et al. have theoretically shown that such a change of Pd-Pd bond length would lead to significant differences on catalysis.[7] In the authors' synthesized Au NBP@Pd nanostructures with varying Pd shell thickness, the surface Pd-Pd bond lengths should be very different with varying Pd shell thickness, and thus strain effect would more significantly influence the H adsorption capacities. However, the authors of this commented paper did not take the strain effect into discussions.

In sum, in the paper "Gold Nanobipyramid-Enhanced Hydrogen Sensing with Plasmon Red Shifts Reaching ≈140 nm at 2 vol% Hydrogen Concentration" by Yip et al.,[1] the statement that electronic effects were the most important effects in their synthesized Au NBP@Pd systems, is unsupported. This is because both ensemble and strain effects should be more significant than the electronic effect in the Au NBP@Pd systems. If there existed Au surface segregation, it is expected that ensemble effects are more important; if there was no surface segregation, strain effect should be more dominant due to the mismatch of the lattice constants between Au and Pd. None of these two situations could support the claim that electronic effect is the most important in this case. Meanwhile, the authors mistakenly cited the paper that has the opposite conclusion against their claim.